\PassOptionsToPackage{unicode}{hyperref}
\PassOptionsToPackage{hyphens}{url}
\documentclass[
]{article}
\usepackage[a4paper, total={6in, 8in}]{geometry}

\usepackage{multicol}
\usepackage{authblk}
\usepackage{amsmath,amssymb}
\usepackage{iftex}
\ifPDFTeX
  \usepackage[T1]{fontenc}
  \usepackage[utf8]{inputenc}
  \usepackage{textcomp} 
\else 
  \usepackage{unicode-math} 
  \defaultfontfeatures{Scale=MatchLowercase}
  \defaultfontfeatures[\rmfamily]{Ligatures=TeX,Scale=1}
\fi
\usepackage{lmodern}
\ifPDFTeX\else
\fi
\IfFileExists{upquote.sty}{\usepackage{upquote}}{}
\IfFileExists{microtype.sty}{
  \usepackage[]{microtype}
  \UseMicrotypeSet[protrusion]{basicmath} 
}{}
\makeatletter
\@ifundefined{KOMAClassName}{
  \IfFileExists{parskip.sty}{%
    \usepackage{parskip}
  }{
    \setlength{\parindent}{0pt}
    \setlength{\parskip}{6pt plus 2pt minus 1pt}}
}{
  \KOMAoptions{parskip=half}}
\makeatother
\usepackage{xcolor}
\usepackage{longtable,booktabs,array}
\usepackage{calc} 
\usepackage{etoolbox}
\makeatletter
\patchcmd\longtable{\par}{\if@noskipsec\mbox{}\fi\par}{}{}
\makeatother
\IfFileExists{footnotehyper.sty}{\usepackage{footnotehyper}}{\usepackage{footnote}}
\makesavenoteenv{longtable}
\usepackage{graphicx}
\makeatletter
\def\maxwidth{\ifdim\Gin@nat@width>\linewidth\linewidth\else\Gin@nat@width\fi}
\def\maxheight{\ifdim\Gin@nat@height>\textheight\textheight\else\Gin@nat@height\fi}
\makeatother
\setkeys{Gin}{width=\maxwidth,height=\maxheight,keepaspectratio}
\makeatletter
\def\fps@figure{htbp}
\makeatother
\usepackage{soul}
\ifLuaTeX
  \usepackage{selnolig}  
\fi
\IfFileExists{bookmark.sty}{\usepackage{bookmark}}{\usepackage{hyperref}}
\IfFileExists{xurl.sty}{\usepackage{xurl}}{} 

\providecommand{\keywords}[1]
{
  \small	
  \textbf{\textit{Keywords---}} #1
}

\title{(Re)framing Built Heritage through the Machinic Gaze}

\author[1]{Vanicka Arora}
\author[2]{Liam Magee}
\author[3]{Luke Munn}
\affil[1]{University of Stirling, United Kingdom \authorcr vanicka.arora@stir.ac.uk}
\affil[2]{Western Sydney University, Australia \authorcr l.magee@westernsydney.edu.au}
\affil[3]{University of Queensland, Australia \authorcr l.munn@uq.edu.au}

\date{October 2023}

\begin{document}
\maketitle

\begin{abstract}

  Built heritage has been both subject and product of a gaze that has been
  sustained through moments of colonial fixation on ruins and monuments,
  technocratic examination and representation, and fetishisation by a
  global tourist industry. We argue that the recent proliferation of
  machine learning and vision technologies create new scopic regimes for
  heritage: storing and retrieving existing images from vast digital
  archives, and further imparting their own distortions upon its visual
  representation. We introduce the term `\emph{machinic gaze}' to
  conceptualise the reconfiguration of heritage representation via AI
  models. To explore how this gaze reframes heritage, we deploy an
  image-text-image pipeline that reads, interprets, and resynthesizes
  images of several UNESCO World Heritage Sites. Employing two concepts
  from media studies -- \emph{heteroscopia} and \emph{anamorphosis} --~we
  describe the reoriented perspective that machine vision systems
  introduce. We propose that the machinic gaze highlights the artifice of
  the human gaze and its underlying assumptions and practices that combine
  to form established notions of heritage.  

\end{abstract}


\pagebreak

\raggedcolumns
\begin{multicols*}{2}

\hypertarget{introduction}{%
\section{Introduction}\label{introduction}}

Built heritage has a long- and well-established relationship with visual
representation and production. Visual regimes of heritage have evolved
and diversified, from carefully curated artistic depictions of the
romantic archaeological ruin to the proliferation of digital photography
on social media produced and consumed in turn by a global tourist
industry. Alongside the far-reaching impact of social media platforms
like Instagram that rely primarily on the production and circulation of
images, sophisticated photo-editing technologies that are no longer
confined to the domain of professionals, have amplified and complicated
styles of aesthetic appreciation of the heritage site. Nevertheless,
across various forms of visual media, the construction of a mythic
representation of the heritage site persists (Dicks, 2000; Watson, 2010;
Sterling, 2016).

Much of the discussion on the construction of heritage identity and
meaning through visual representation or the `heritage gaze' has focused
on two closely entwined ways of seeing: the tourist gaze (Urry, 2002
{[}1990{]}; Urry and Larsen, 2011; MacCannell, 1999; Watson, 2010,
Waterton, 2009) and the expert gaze (Smith, 2006; Winter, 2006;
Moshenska, 2013). In the context of archaeological monuments and sites,
both ways have both been further tied to forms of mechanical
apprehension and capture since the inception of photography. Maxime Du
Camp's French government-commissioned images of Egyptian monuments
exemplify the early enthusiasm for architectural and landscape
photography in the years following the invention of the daguerreotype,
surpassed only by the desire to reproduce the human form (Kittler, 2010;
Urry and Larsen, 2011). Photography has long assisted in what Sterling
(2019: 2) has termed the `mythic representation of heritage as
ideology', drawing attention to iconic or emblematic aspects of sites
that reinforce narratives of power. Analysis of the performative
function of photography has often been connected to the concept of the
`gaze': the work conducted by the camera-and-operator to compose and
constitute the heritage subject. For scholars such as Urry and Larsen,
the act of seeing is always informed by `ideas, skills, desires and
expectations' (2011: 1), and in the era of mass tourism -- since the
nineteenth century, and so coinciding with the era of mass photography
-- the seeing of sites has been especially demarcated by the `tourist
gaze.'

The emergence of computer vision and, recently, of machine learning
systems trained on image corpora reproduce both these modes alongside
other styles and subjects, retaining as they do so existing social
biases (Offert and Phan, 2022). However, this reproduction is not pure.
In their reconstitution of heritage, image-generating systems such as
Midjourney observe conventions with palettes and perspectives, but also
at times inject the uncanny differences of an alien observer or subject
(Parisi, 2019). As these systems scale and mature, it no longer seems
sufficient to ascribe these differences to the category of `error' -- an
intermediate stage in a system's evolution, to be `fixed' with the next
software update. Differences between their outputs and human expectation
seem to belong instead to a novel \emph{sui generis} mode of visual
perception and production, which we describe here as the `machinic
gaze.' By directing computer vision algorithms to interpret and
resynthesise a controlled set of images of heritage sites, we attempt to
register aspects of this gaze. In doing so, we ask the question: What
does the machine see when it looks at heritage?

A feature of this `gaze' is its intensification. The proliferation of
social media, and Instagram in particular, has stretched and magnified
how it has become invested and circulated within the circuits of travel
capitalism (Barauah, 2017; Ogden, 2021; Oh, 2022). This expansive
digitisation of vision has led in turn to new possibilities in how
machines consume and produce images. Social media image agglomerations
have been systematised and organised into vast archives like LAION
(Schuhmann et al., 2022), a data store of five billion image-caption
pairs that in turn has been instrumental in the training of
Vision-Language Models (Zhang et al., 2023). With respect to these
systems, two distinct kinds can be distinguished: \emph{image-to-text}
auto-captioning systems such as BLIP-2 (Li et al., 2023), and
\emph{text-to-image} generative systems such as Stable Diffusion,
Midjourney and Dall-E (Mostaque, 2022; Midjourney, 2022; Ramesh et
al., 2022).

It is the second type of system that we focus upon especially here,
though as discussed in the methods section, we examine examples of the
first as well. With generative AI text-to-image systems, the input of a
text `prompt', an instruction made up of typically English words that
specify a subject, style and format, generates synthetic images that,
despite having no direct referent in their training sets, can integrate
parts of that prompt in often evocative and striking ways. This
synthesis is possible via an intensive process that uses a previously
trained neural network to progressively modify what is at first a random
image into one shaped by the set of tokens included in the prompt. Both
prior network training and this process of image synthesis work to
produce recognisable pictorial representations through purely
computational processes. This recognisability is at the same time
thoroughly disjoined from the `real' that Crary (1992), for example,
identifies as embedded in modes of human vision and perception.
Accordingly, this way of seeing simultaneously constitutes a culturally
and commercially generative system of visual apprehension, and a `gaze'
that is askew and uncanny (Parisi, 2019).

We focus here on how this apprehension works to reproduce visual
representations of heritage sites that have been subject to the explicit
focus of the heritage gaze. Specifically, we selected a small group of
images from UNESCO's World Heritage database of sites. We developed a
small Python application that first uses a mix of technical methods to
read and decode these images into textual prompts, and then submits
those prompts to AI image-text systems to generate candidate reimagining
of the original images. Our purpose here was not to evaluate the
fidelity of either auto-captioning or image generation systems. Instead,
it was to produce a small archive of internally diverse representations
that demonstrate semantic patterns and structures that emerge in
contemporary mechanised ways of seeing. These enabled in turn a
commentary on the relationships between the generated image and its
prompt, the represented object of heritage and -- more speculatively
--~its human interpreter.

We undertook this exercise with two objectives. The first is to consider
the ways in which the image model captures, `understands,' and recreates
the heritage site and the specific gaze that is directed towards these
sites. The second is to extend the exploration of the politics of visual
representation of global sites of heritage through the medium of the
synthetically produced image. Properties of this synthesis, we argue,
can condense, and refract highly disparate human representations of
heritage, marking out more clearly its own preoccupations and
ideological attachments.

\hypertarget{conceptualising-the-machinic-gaze}{%
\section{Conceptualising the Machinic
Gaze}\label{conceptualising-the-machinic-gaze}}

The gaze, often with attached qualifiers (`male,' `colonial,'
`tourist'), has an extensive history in heritage and adjacent fields of
cultural studies. A common thread to distinct conceptualisations, from
Mulvey's (1975) seminal essay on the male gaze to Urry and Larsen's
(2011) discussion of the tourist gaze, is that \emph{seeing} is never
only a perceptual act, but is always informed by background assumptions,
desires, prejudices and power relations that inform interpretation of
what is seen. As Urry and Larsen further analyse, such background
considerations apply no less when the gaze is directed toward nonhuman
artefacts: landscapes, buildings, exhibits, and other cultural artefacts
packaged for consumption by the tourist. This interpretation is
re-doubled by the profound effects of photography and social media,
which reify and circulate images of such objects without in any way
diluting consumer desire to attend to the original. However, no gaze
ever consumes its object fully. Waterton and Watson (2015: 26) for
example describe another mode of apprehending visual heritage, one which
is configured and curated by experts `around a nexus of value endowed by
pastness, scarcity and aesthetics.' Even when it rubs up against the
tourist gaze, in the form of the expert guide for instance, it remains
differentiated in its attention to masterly knowledge and `official'
narratives of heritage making' rather than the mundanely picturesque as
its primary register (Sterling, 2016). And even critics of the tourist
industry acknowledge how the non-expert viewer or tourist is not merely
a passive spectator but participates in the construction of meaning and
performance of heritage places (Urry, 2004; Smith 2006). The tourist
gaze is polyvalent and complex (Urry and Larsen, 2011) and as Sterling
has argued (2016: 248) in the context of heritage, itself only one
element in a network that also must encompass globalisation, capitalism,
technical practices of photography and the individualised affective
engagement with place.

Common to many discussions of these two systems -- the tourist and the
expert gaze -- is attention towards their shared outwardly-directed
nature. In both cases, the gaze is conferred \emph{by} an individual or
collective human subject \emph{towards} an artefact or site, an act
which serves to constitute or enmesh the object within various
discursive and interpretative frames. MacCannell's (2001) has critiqued
Urry's unidirectional analysis of the gaze -- borrowed from Foucault's
analysis of the medical field's interested observation of the patient's
body -- with an alternative account drawn from Lacanian analysis. Though
Lacan's (1977) famous Seminar XI treatment oscillates between two
meanings of the French \emph{le regard} -- translated into English as
the \emph{look} and the \emph{gaze} -- it is the second of these
meanings, both counterintuitive and influential, that concerns
MacCannell. According to Lacan the gaze is what takes place when the
human observing subject becomes aware that they \emph{themselves} have
entered the frame of observation. No longer objective, the viewer
becomes a subject precisely at this moment of uncanny recognition --
when they catch themselves in a mirror, or less directly, become aware
of their situatedness as an observer.

MacCannell's application of this reflexive sense of the gaze helps, in
his argument, to recuperate the agency of the heritage spectator.
Viewing heritage does not simply involve a consuming tourist or
calculative state, but -- at least at particular moments --~involves a
transformation of the spectator into a subject aware of their own
historicity. The tourist experiences for example the strange sense of
becoming an object for some other, future viewer or visitor -- and as
this object, also becomes a proper subject. Resisting efforts to subsume
all touristic appreciation to that of cliche, Sterling (2016) has
similarly argued that the seeing tourist is also an embodied figure, one
who apprehends their own materiality in heritage encounters, and to
varying degrees is also managed through deliberately arranged scaffolds
and signs by heritage site managers. The body in Sterling's account in a
certain sense anchors the otherwise cliched gaze within the singularity
of the individual subject.

Despite for example Urry and Larsen's interests in photography and
social media, these varied articulations share the sense that the holder
of the gaze is human. The development of visual language models that do
not capture or store images constructed by a human operator seems to
expand how `gaze' must be conceptualised. In describing how machines see
with and on behalf of humans, Denicolai (2021) invokes various terms,
including `machinic vision', `mechanical perception' and `robotical
gaze' -- expressions we collapse into the term `machinic gaze'. How this
gaze perceives objects and sites of heritage reflects, in part, the
regard of human engineers and photographers who design and feed data
into algorithms. We argue however that this synthetic reflection is not
a simple transparent mirroring of a collective and diffused human
apprehension of heritage. More is involved, in other words, than a
mechanical reproduction of the touristic, expert, or other forms of
human seeing. The technical perception and production of images involves
instead a different and inhuman gaze that is directed towards heritage,
creating associations of form, aesthetics, and value, and constructing a
reflected heritage myth. In the context of machine learning, Mackenzie
and Munster (2019) have described the general synthetic production of
images via algorithms as a kind of `invisuality', a neologism that
conveys their inward-looking quality. Rather than responding to a set of
natural or cultural conditions -- in the sense that Urry describes the
`tourist gaze' for instance -- image synthesis relies exclusively upon
the intricate mappings of language terms and clusters of pixels
established through the system's training. Image synthesis constitutes
an inhuman kind of `platform seeing' (Mackenzie and Munster, 2019), with
its outputs constitutive of a concentrated yet heterogenous form of
`mechanical reproduction' (Benjamin, 1986).

This machinic gaze can be further distinguished by the precise
operations performed by the technical apparatus: observing via a camera,
analysing an image, compositing an image from trained patterns, or some
combination of these actions. In each case the machinic gaze carries
forward both the direct aesthetic meaning of an apprehension of an
object, and the indirect or reflexive meaning. This latter sense
requires an adjustment, a three-way relationship between the object, the
machine, and the human subject who then looks upon the machine's
synthetic representation of the object, observing in it -- in its
similarities and differences to how we ourselves might interpret the
object -- a form of refracted mirroring of the human subject itself. In
looking at heritage artefacts, this in fact performs a double mirroring,
as the object of heritage already contains within itself a form of
reflection of the observer. Following MacCannell's discussion of Urry's
notion of the gaze, we expand upon this reflexive property in terms of a
reversal of lens back upon the spectating and contributing human
subject.

\hypertarget{methods}{%
\section{Methods}\label{methods}}

In this section we map this conceptual understanding of the machinic
gaze to a small set of images that were analysed and used to generate,
through algorithms, other, purely synthetic images. These activities
were conducted in sequence: we selected digital photographs, then
analysed these images with three procedures (BLIP-2, Google Vision API,
image EXIF metadata) to assemble a brief textual description for each
image. These assembled descriptions were then submitted in turn to three
image generation systems in the form of `prompts' (Stable Diffusion;
Realistic Vision, and MidJourney) to produce a series of image samples
-- 240 in total. Finally, we interpreted these images in terms of
subject, style, composition, and deviations from the source images. We
briefly discuss each of these steps below.

\hypertarget{i.-image-selection}{%
\subsubsection{i. Image selection}\label{i.-image-selection}}

We used the archive of photographs from the official website of UNESCO's
World Heritage Sites as base images. Most of these photographs have been
taken by experts appointed directly by UNESCO's World Heritage Programme
Office or by individual State Parties and are intended to serve
simultaneously as official visual documentation of the site and
ostensibly communicate a sense of its `outstanding universal value.' We
filtered sites on the UNESCO World Heritage in Danger list and limited
our selection to sites that were cultural and those which met criterion
(iv), `to be an outstanding example of a type of building, architectural
or technological ensemble or landscape which illustrates (a) significant
stage(s) in human history' (UNESCO, 2008). These selection criteria
ensured a consistency in the vocabulary of the ways the sites were
documented and presented.

Our final selection of the five site images (see Appendix) was motivated
by several factors. First, we wanted to work with a limited set of
images that had a common underlying subject, but which are marked by
visibly contrastive features. Second, we wanted buildings that are
distinctive in form but likely underrepresented in the LAION training
set -- `iconic' heritage sites such as the Eiffel Tower would likely be
oversampled and therefore more readily `reconstructed' by text-to-image
models. Third, we selected images that highlighted both contrastive and
canonical or representative features of the sites. Having selected sites
based on a set of common properties, we then selected photographs of
buildings or urban settings that instead contrasted with respect to
building typology, geographic region, historical style, and the
description of function of the site. We generally opted for photos that
were medium range (which were in the majority), as close-ups and distant
shots obscured specific cultural and geographic features. In one case
(Sana'a), we chose instead a wide-angle cityscape. In all cases, we
selected photographs of intact buildings rather than ruins, again to
maximise opportunity for the algorithms to identify distinguishing
features. We also only selected Creative Commons-licensed photos.

\hypertarget{ii.-expert-and-machinic-readings-of-heritage-sites}{%
\subsubsection{ii. Expert and machinic readings of heritage
sites}\label{ii.-expert-and-machinic-readings-of-heritage-sites}}

For each of these five sites, we employed two techniques to `engineer'
prompts. The first reflects the `expert' view of the site, taking the
UNESCO-supplied description as the verbatim prompt. The second used a
series of technical approaches to extract information from the selected
site photos, and then constructed a synthetic and entirely automated
prompt. In producing the two types of prompts we compare how expert and
machinic readings or `gazes' of these sites produce distinct (or
similar) aesthetic visual results.

Technique one is rooted in geographical aspects of place and in
`qualitative' or value-based descriptions of place. The UNESCO WHS
descriptions are designed to evoke a sense of place in a way that
mirrors the accompanying photographs. The photographs we selected have
been taken by UNESCO-appointed experts and for the most part a similar
group of experts write the descriptions of the sites. Both privilege
certain kinds of expert `viewing' or a gaze of the sites themselves,
attentive to what is most salient, distinguishing or `outstanding,' and
what therefore has to be compared against a repertoire of other built
forms and sites. We used the UNESCO WHS \emph{Description} tab as the
text from which to select and edit the prompt. Our decision to focus on
criteria (iv) sites, allowed us to closely examine UNESCO's own emphasis
on visuality and aesthetics as constitutive of heritage
value.\footnote{This technique has limitations: sites which have been
  nominated before the UNESCO Operational Guidelines have been ratified
  have had much shorter and less `rigorous' nomination dossiers which
  meant shorter and less detailed `description' sections. Some of these
  descriptions have not been updated since the site was first nominated.
  Some photographic galleries were much smaller and contained older
  images that may not accurately reflect the state of the site today.}

To consider how algorithms `see' these images, we employed a mix of
algorithmic and deep learning techniques, ranging from extracting
captions via BLIP-2, to probabilistic analysis of images via Google
Vision's API and extraction of image EXIF metadata. Each of these modes
of machinic reading, produced a textual description. BLIP-2 is a
vision-language based model which uses pre-trained image encoders and
large language models to extract textual descriptions of images through
an encoder-decoder transformer system (Li, Li, Savarese, Hoi 2023). Here
we use BLIP-2 to generate short textual descriptions and other
text-based queries on our base image set.

To enrich the prompt, we combined this text with labels extracted from
Google Vision's Application Programming Interface (API). These labels
included computed image properties such as dominant colours, objects,
locations, architectural features, geometric shapes, natural features,
colour schema, as well as individual objects. This machinic reading was
able to arrive at probabilistic responses that identified certain
building categories and adjectives, like `monument' and `heritage.'
Equally, it was inexact and imperfect; often identifying features of the
place, like geographical locations, precise architectural styles, and
historic periods, were omitted or misidentified. We also extracted
metadata from the digital image itself, including the type of camera,
focal length, exposure time, and use of flash. These parameters helped
in building the final prompt when we attempted to resynthesise the image
content.

\hypertarget{iii.-machinic-synthesis-and-its-interpretation}{%
\subsubsection{iii. Machinic synthesis and its
interpretation}\label{iii.-machinic-synthesis-and-its-interpretation}}

The prompts or instructions produced through both methods were then
submitted to three image generating systems. We selected Stable
Diffusion and Midjourney since in 2023 these were the most widely
discussed text-to-image systems, eclipsing the earlier attention paid to
OpenAI's Dall-E model. Stable Diffusion is a text-to-image model that
has been made open source by its developer, StabilityAI, and can be
downloaded and operated on consumer devices. Midjourney is a
service-based system that requires a subscription to operate, via the
Discord social media platform. Both perform similar functions,
converting a natural language prompt into one or more images that aim to
`represent' that prompt in a meaningful way. We used the latest versions
of these two systems at time of writing: version XL in the case of
Stable Diffusion and version 5.2 in the case of Midjourney. As Stable
Diffusion models are released as open source, they can be adapted, or
`fine-tuned,' on much smaller data sets of images to produce particular
styles or aesthetics. For further contrast we used an older version of
Stable Diffusion (version 1.5) that had been fine-tuned to generate
photorealistic images, in a model named `Realistic Vision.'

For each of the five sites, we applied the prompts generated through the
approach described above to each of the three systems. We specified each
system to generate four images for each prompt, producing a data set of
120 images (5 sites x 3 systems x 4 images). For comparison, we also
applied the UNESCO-supplied description for the selected site as a
prompt to the same combination of site, system and image variations,
doubling the size of our data set to 240.\footnote{To aid reproduction,
  and to support similar studies, the scripts and datasets have been
  published on GitHub.} Together these stages -- source image selection,
image analysis, image generation, image interpretation -- conducted
across different technical embodiments enable a characterisation of the
machinic gaze of heritage sites.

Finally, we interpreted these sets of images in terms of their
composition, form, subject selection, framing, and colour palette. This
interpretation, as we reflect upon in our findings and discussion,
involves reflection upon the acts of seeing and reading of
machine-generated images. It builds necessarily upon our own backgrounds
in heritage and media studies, and consequently involves a specific form
of what has been theorised as the expert gaze. Despite the obvious
limits of such interpretation, we look to avoid a specific judgement
upon these machinic productions in terms of their approximation to some
notion of `ground truth' or as a quantitative exploration of bias within
the underlying datasets of these systems (Salvaggio, 2022).

\hypertarget{findings}{%
\section{The Machine Imagines Heritage}\label{findings}}

We discuss here three sets of images that contrast internally (across
models and prompt) and externally (across sites). We use `H-M' and `M-M'
to distinguish human-prompt-machine-generated from
machine-prompt-machine-generated images.

\hypertarget{old-towns-of-djennuxe9}{%
\subsubsection{Old Towns of Djenné}\label{old-towns-of-djennuxe9}}

\emph{Figure 1} shows a mid-distance elevational aspect of the mosque of
Djenné, which is one the key structures identified in the description of
the World Heritage Site. The adobe mosque appears in multiple
photographs of the site, as one of the distinctive architectural
landmarks within the urban ensemble of Djenné. The photograph frames the
mosque tightly, editing out the immediate context of the marketplace or
townscape that surrounds the mosque.

\begin{figure*}
  \centering
  \includegraphics[width=4.4579in,height=3.33891in]{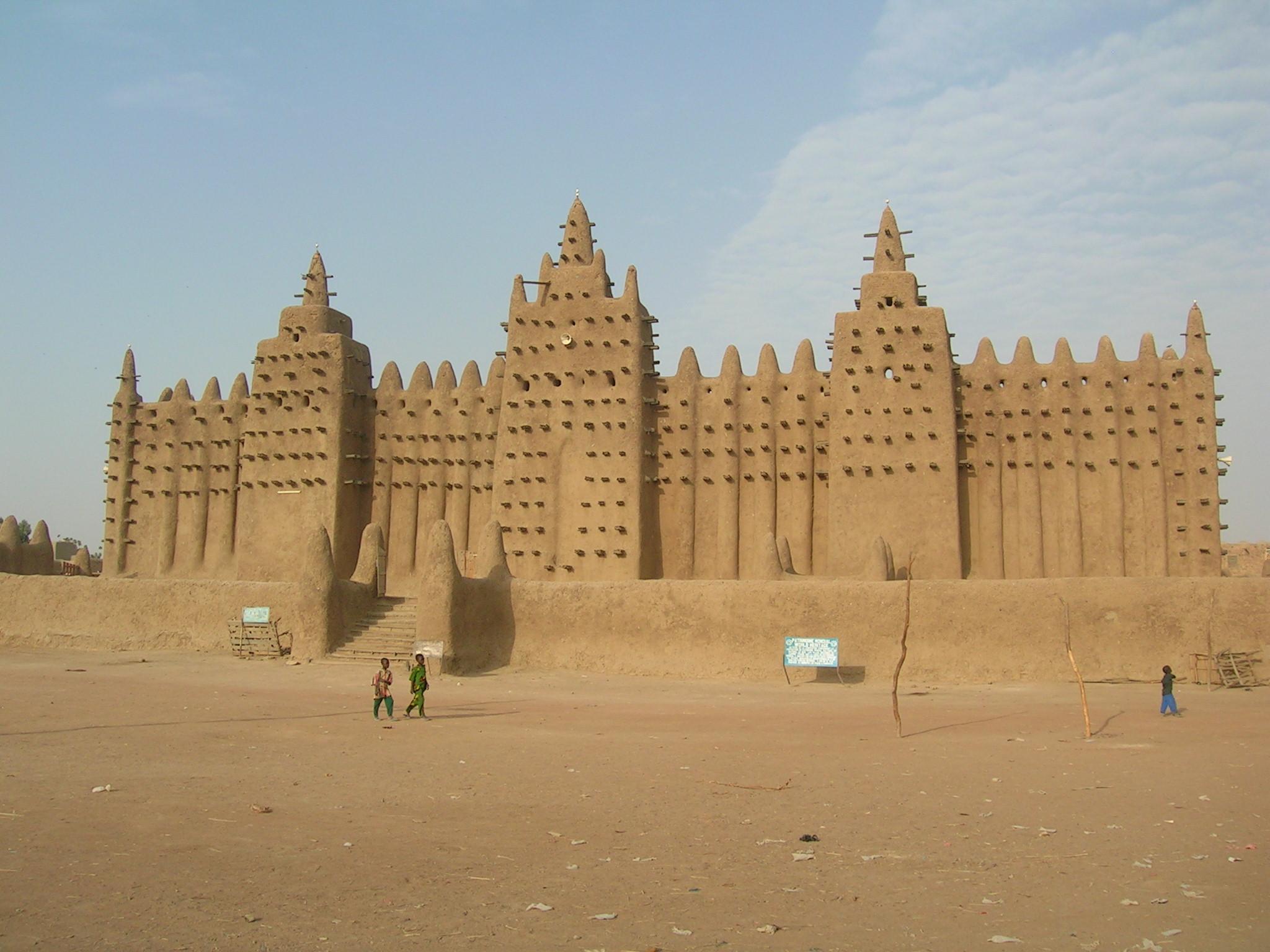}
  \label{fig:fig1}
  \caption{Old Towns of Djenné Source: UNESCO}
\end{figure*}

\begin{figure*}
  \centering
  \includegraphics[width=5.3641in,height=5.45592in]{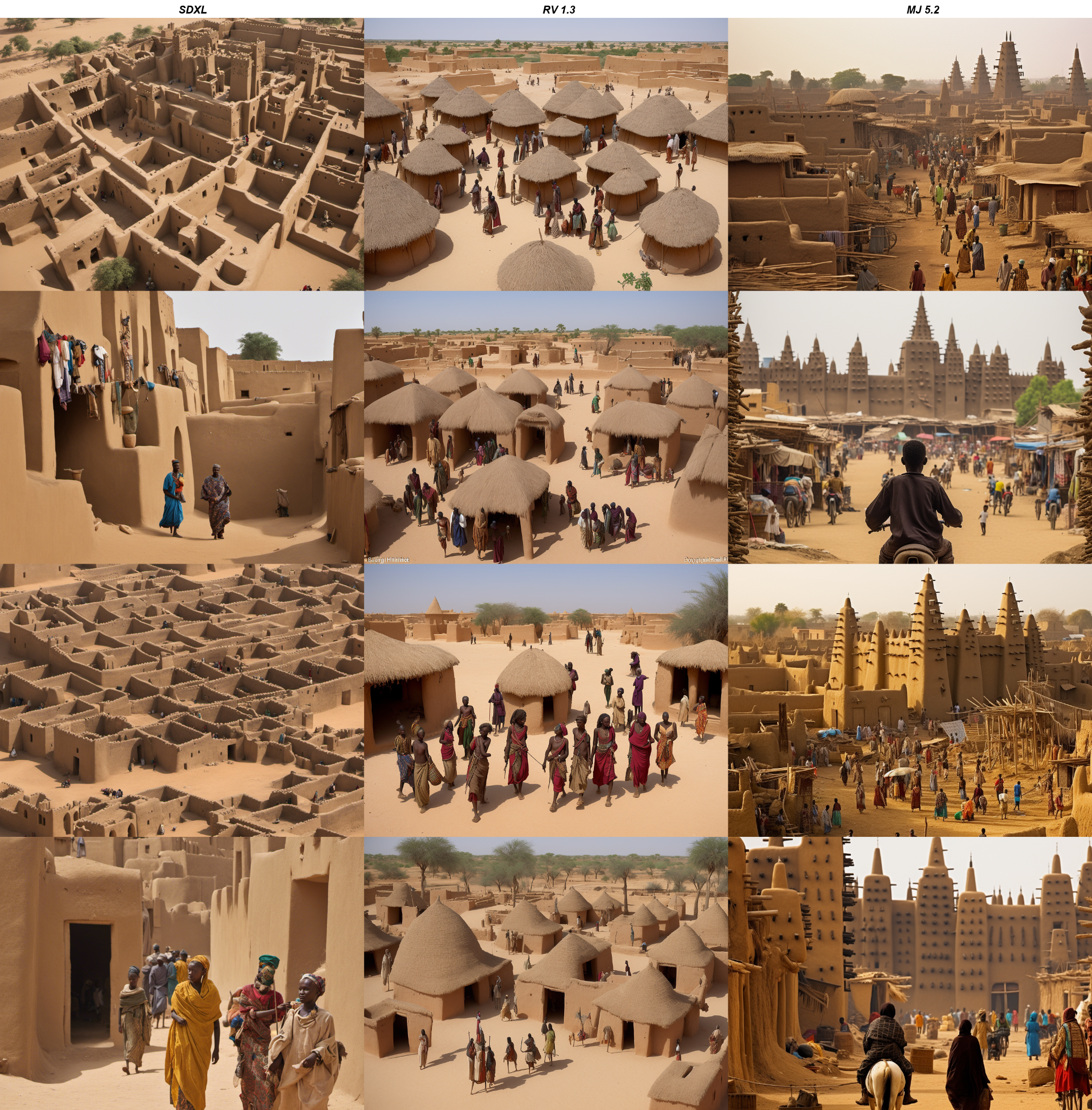}
  \label{fig:fig2}
  \caption{Human Prompt / Machine Generation; SDXL, Realistic
  Vision, Midjourney}
\end{figure*}

\emph{Figure 2} shows results of the `H-M' process: four outputs (in
rows) of three image models (in columns), in response to the prompt that
was extracted from the description of Djenné on the UNESCO website,
which included phrases like `typical African city', `intensive and
remarkable use of earth', `mosque of great monumental and religious
value' (UNESCO, 2023). In the case of Stable Diffusion (both versions),
while the colour scheme of the UNESCO image is retained, no version of
the mosque is produced in any of the images. SDXL (left column) shows,
at different resolutions, a grid-like configuration of mud brick
structures that approximates the sub-Saharan vernacular, but without the
specificities of Djenné's architectural proportions or ornamentation.
The Realistic Vision outputs (centre column) produce an approximation of
a generic sub-Saharan settlement, small adobe buildings with thatched
roofs --~neither characteristic of the mosque nor of the general town.
Midjourney (right column) produces images that are quite distinct from
the reference image, but that resemble other images of the townscape of
Djenné, showing markets, houses, and people in transit. This set of
images shows in particular the compositional nature of Midjourney's
generated images: in each case some version of the mosque is
recognisable, but in the background, shot in shadow and occasionally at
oblique angles. People in the foreground feature in a quasi-cinematic
way: in two cases, one or two people appear close to the presumed
camera, as though on a journey, while more distant figures appear as
accidental subjects. In all cases, people appear in some variant of an
assumed local dress.

\begin{figure*}
  \centering
  \includegraphics[width=5.28486in,height=5.37532in]{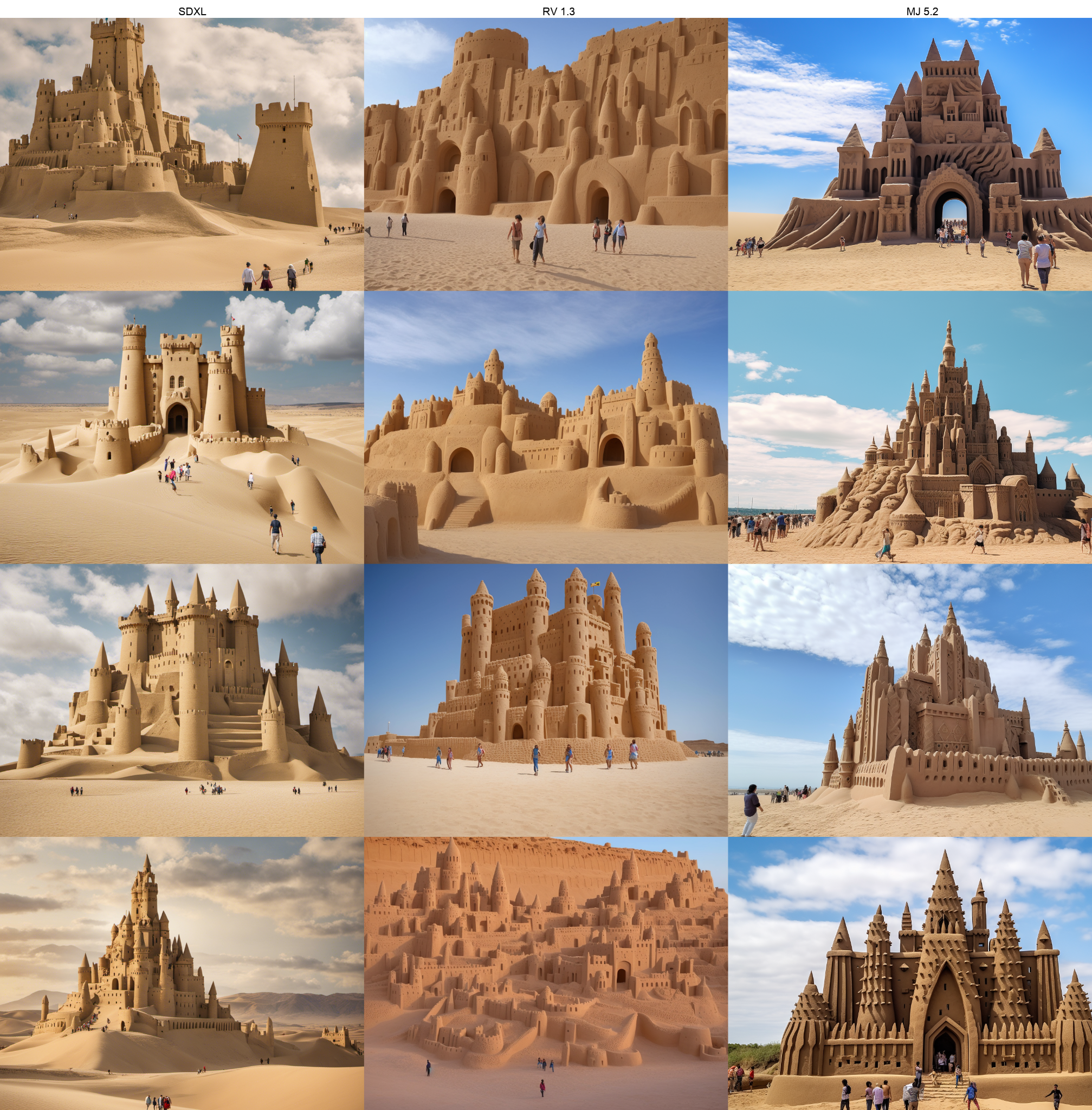}
  \label{fig:fig3}
  \caption{Machine Prompt / Machine Generation; SDXL, Realistic
  Vision, Midjourney}
\end{figure*}

For the `M-M' reading of the source image, we obtained the following:

\begin{quote}
\emph{photograph of a large sand castle with people walking in front of
it}, \textbf{Building center, Sky, Cloud, Travel, Landscape, Sand,
Aeolian landform, Facade, History, Ancient history, Archaeological site,
Historic site, Art, Arch, Soil, Horizon, Singing sand, Tourism, Castle,
Desert, Tourist attraction. Colors: \#bb9667, \#b7c1c3, \#b78e5c,
\#9b7344, \#977649, \#6f4b1f, \#aa9373, \#694e26, \#634d2d, \#8d795a}
Shot with a E3700, at a resolution of 300 pixels per inch, year 2005,
exposure time of 5/1806, Flash did not fire, auto mode, focal length of
27/5
\end{quote}

The first part, in italics, represents the BLIP2 caption; the second, in
bold, a textual representation of properties extracted from the Google
analysis; and the third, metadata properties of the source image. The
misrecognition of the mosque as a sandcastle in the machinic reading can
be attributed to an alignment of the language of castles with similar
visual patterns in the training data. \emph{Figure 3} represents the
outputs, following the same pattern as \emph{Figure 2}.

In each case, the anchoring characteristic is the first part of the
prompt, `large sandcastle.' In one case (Midjourney, bottom), there are
recognisable aspects of the source image, including the mosque's
exterior ornamentation. But in most cases the `castle' produced more
closely resembles Disneyfied castle caricature, diverting quite starkly
from the rectilinear form of the mosque. A recurring similarity in most
of the images produced is a lack of surrounding built context, both the
mosque, and the castle appear to be isolated monumental objects in the
frame.

In other respects, and despite the prompt specifying colours, camera
type, and exposure time, the reference to sandcastle appears to
over-determine the colour palette and saturation level. Compared to
Figure 2 (H-M), in Figure 3 (M-M) the sand, of both castle and
foreground, is lighter, and the sky clear rather than hazy. Keywords
such as `history' and `archaeology' also change the sense of scale and
context, with the implied camera position being now more distant. The
scene is also deracinated: the form of the `castle' is drawn from a wide
range of typological and stylistic references, and though diminutive,
the `people' referenced in the prompt are dressed in global rather than
`local' attire, tourists who apprehend the monumental structure rather
than locals who live around it. The presumed holder of the gaze is, in
other words, no longer solely a figure imagined as behind the camera,
but firmly embedded within it.

\hypertarget{old-city-of-sanaa}{%
\subsubsection{Old City of Sana\textquotesingle a}\label{old-city-of-sanaa}}

\emph{Figure 4} shows the original UNESCO image (top) of the Old City of
Sana'a, along with two generated outputs from Midjourney 5.2: the first
(middle, human-machine or ``H-M'') is the result of the UNESCO,
human-authored description used as a prompt, and the second (bottom,
machine-machine or ``M-M'') the output of the machine-generated prompt.
In this case, the UNESCO description places emphasis on the cityscape,
with phrases like `rammed earth and burnt brick towers,' `densely packed
houses, mosques' but also specifies colours -- `white gypsum' `bistre
colored earth,' `green bustans.' The BLIP generated prompt correctly
identifies the image subject -- `old city of Yemen' but also the frame
-- `an aerial view.'

\begin{figure*}
  \centering
  \includegraphics[width=3.80161in,height=7.66107in]{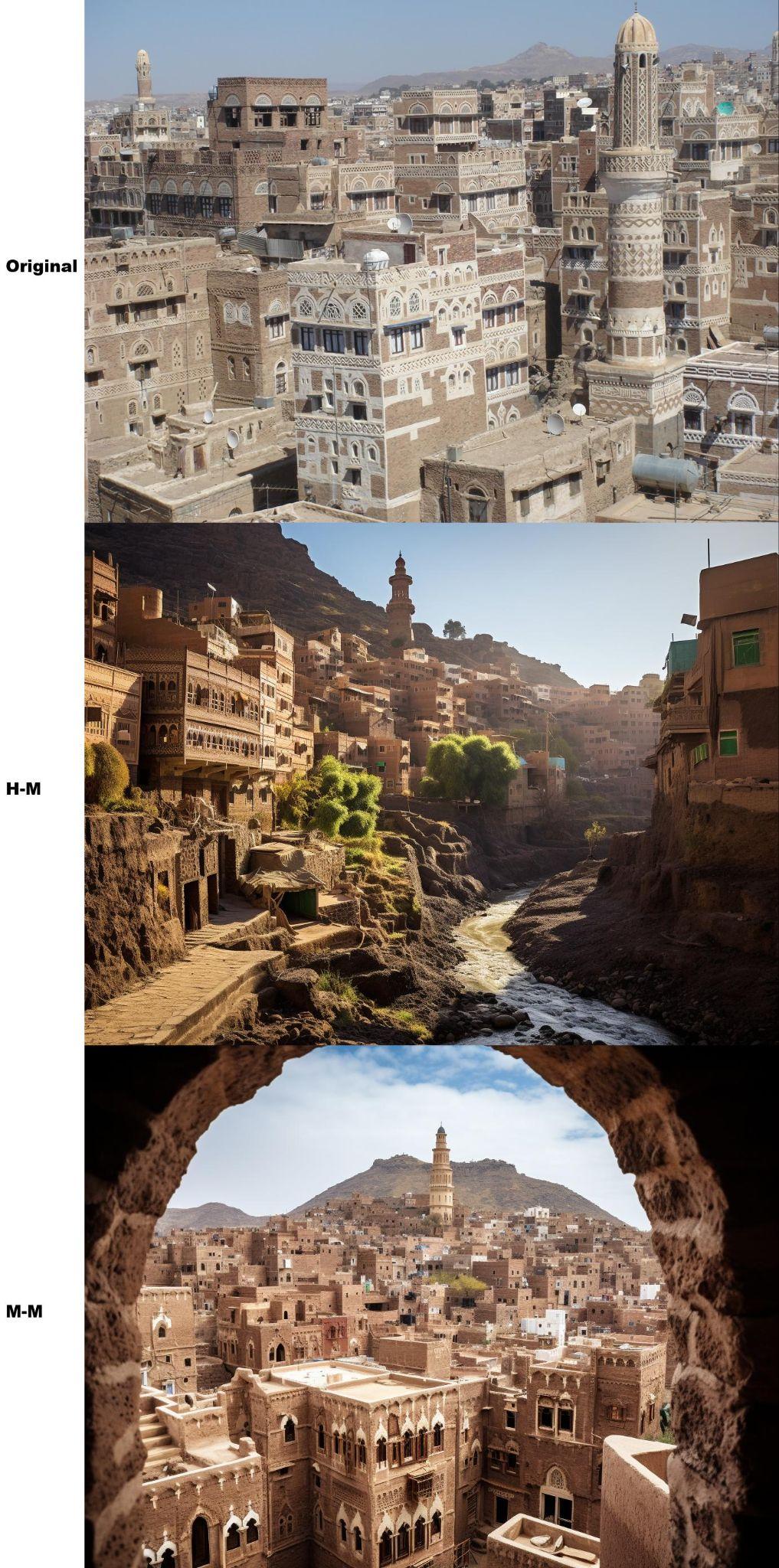}
  \label{fig:fig4}
  \caption{Sana'a; Top: Original; Middle: Midjourney (H-M); Bottom: Midjourney (M-M)}
\end{figure*}

Here the machine-generated prompt was:

\begin{quote}
photograph of an aerial view of the old city of yemen, \textbf{Building
center, Sky, Daytime, Window, Architecture, Landscape, City, Urban
design, Landmark, Cityscape, Facade, Roof, Human settlement, Urban area,
Medieval architecture, Metropolis, Arch, Mixed-use, Archaeological site,
Ancient history, Historic site, History, Turret, Dome, Town, Monument,
Bird\textquotesingle s-eye view, Tourism, Classical architecture, Holy
places. Colors: \#cdc3b8, \#cfc2b1, \#ab9b8a, \#a69b93, \#e7ddd2,
\#83766e, \#887868, \#e9dccb, \#5f534d, \#3f342e.} \emph{Shot with a
DSC-T9, at a resolution of 72 pixels per inch, year 2009, exposure time
of 1/500, no flash, focal length of 1139/100}
\end{quote}

As was with the case of the Midjourney outputs for Djenné, both
generated outputs show a tendency to emphasise geographic features
identified in the textual description or source image. Mountains are
exaggerated; and in the `H-M' case parts of the city hug a cliff-face
and overlook a river, in sharp contrast to the source photo. Tonally the
`H-M' image also employs stronger use of contrast (brightly illuminated
buildings on the left compared to those in shadow on the right), and a
greater colour dynamic -- browns, vivid blues, and varying greens --
reflects the especially chromatic verbal description (`spacious green
bustans').

The `M-M' image on the other hand is strikingly similar, both in broad
elements of architectural form and image composition, to the source.
Just as with `giant sandcastle' in the case of Djenné, here both the
identification of aspect (`aerial view') and location (`old city of
yemen') work to determine scale, perspective and chromatism of images
for all three models. In the case of the selected Midjourney image, the
identification of an `arch' object by the Google API -- barely
discernible in the source image --~is brought into the fore as a
photographic conceit, as `found' frame for the distant cityscape. Though
not evident in this source image, even another of the UNESCO images of
Sana'a employs the same framing device -- a convention of the `serious'
or expert photographer the machine has learned to reproduce. Despite the
inclusion of a palette extracted from the source though, the colours of
the sky and buildings are once again more lurid and saturated than those
that appear in the official `expert' gaze -- a kind of machinic
equivalent to an Instagram filter designed to appeal instead to some
imagined would-be tourist to the city.

This last feature is unsurprising for several reasons: the training sets
include more `tourist' than `expert' images, reinforced by the very
inclusion of the term `tourism', alongside `archaeological history' in
the generated prompt; more contemporary images featured in those
training sets also use greater colour range than even those from the
2000s decade; the reference to a specific location; and finally,
Midjourney itself is a commercial system that has been `fine-tuned' to
produce arresting images precisely through use of high contrast. And
yet, in the final case we discuss here, this effect is in fact reversed.

\hypertarget{tombs-of-buganda-kings-at-kasubi}{%
\subsubsection{Tombs of Buganda Kings at
Kasubi}\label{tombs-of-buganda-kings-at-kasubi}}

\emph{Figure 5}, featuring representations of the Tombs of the Buganda
Kings at Kasubi, uses the same pattern as \emph{Figure 4}: at the top is
the original UNESCO image, followed by two synthetic images, this time
generated by Stable Diffusion XL, selected for the purpose of contrast.
The middle image (H-M) is again produced from the UNESCO textual prompt,
while the bottom image (M-M), from a prompt constructed from
machine-generated captions and image metadata. The UNESCO description in
this case emphasises the materiality of structures, with the phrases
`organic materials,' `wood, thatch, reed, wattle, and daub' but also
references form `circular and surmounted by a dome.' The machinic prompt
locate the structure and identifies the image as a `photograph of the
roof (is) made of straw.'

\begin{figure*}
  \centering
  \includegraphics[width=3.85706in,height=7.56775in]{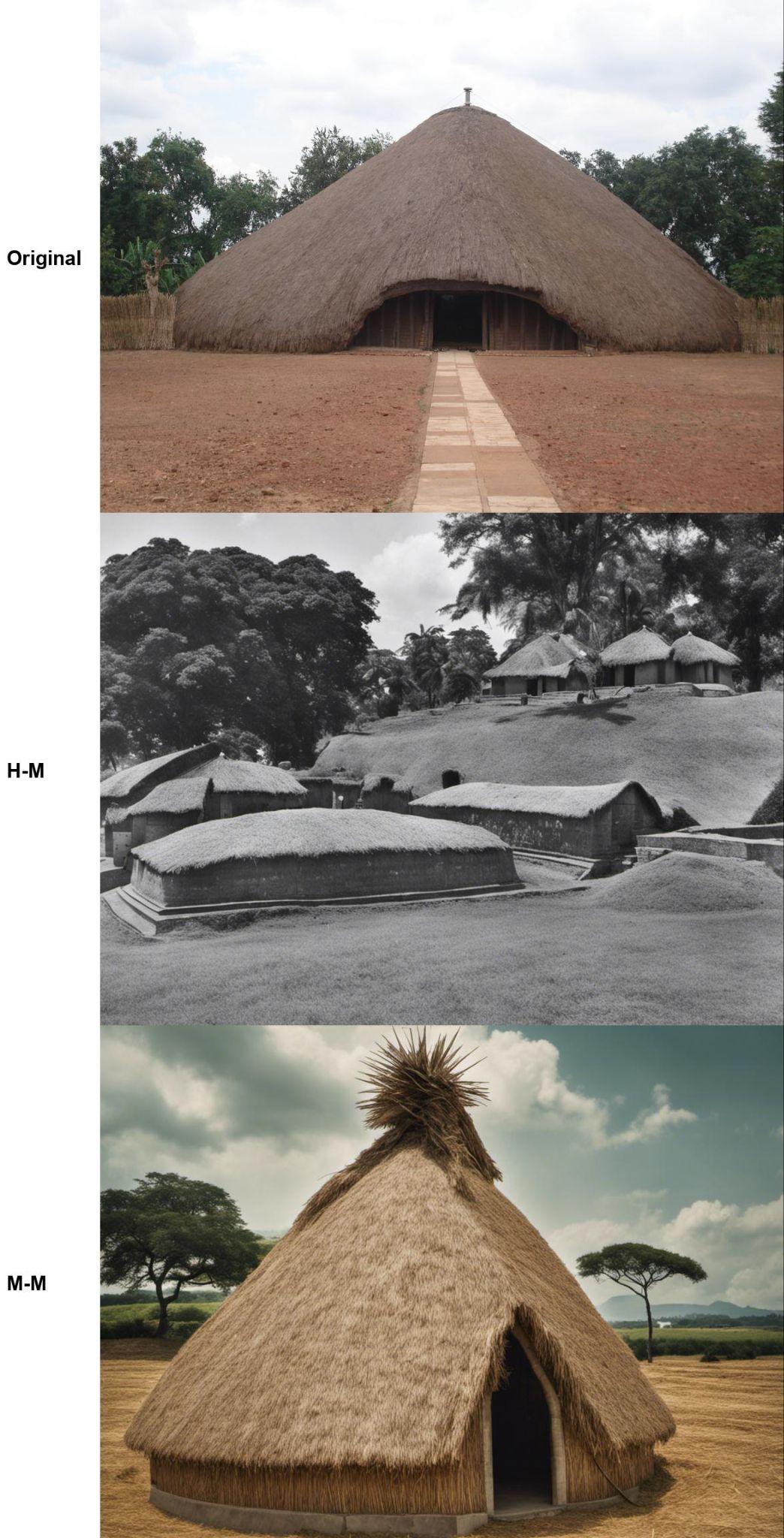}
  \label{fig:fig5}
  \caption{Kasubi; Top: UNESCO; Middle: SDXL (H-M); Bottom: SDXL (M-M)}
\end{figure*}

Machine-generated prompt:

\begin{quote}
photograph of the roof is made of straw, \textbf{Building center, Cloud,
Sky, Land lot, Tree, Thatching, Shade, Grass, Tints and shades, Roof,
Monument, Triangle, Soil, Historic site, Symmetry, Landscape, Building
material, Hut, House. Colors: \#83726b, \#9a7360, \#392d29, \#d7b9a4,
\#f2f3f6, \#7c685d, \#211918, \#bb9783, \#645650, \#6b584b.} \emph{Shot
with a DSC-W50, at a resolution of 72 pixels per inch, year 2007,
exposure time of 1/80, Flash did not fire, auto mode, focal length of
47/5}
\end{quote}

The first photograph of the Kasubi tombs, representing a front
elevational aspect to the main structure, focuses primarily on the
symmetry of the structure, its materiality (\textquotesingle thatch and
reed' in particular) and form, while the tight framing of the camera
angle and the relative absence of other objects and context to add a
sense of scale, creating a sense of monumentality of the fairly austere
structure. The photograph of the single structure devoid of context
emphasises a monumentality that is not reflected in the UNESCO
description, which instead identifies intangible aspects of the tomb
including the continuity of its use and its associated meaning. These
non-visual cues acknowledge that the building aesthetics and form are
not solely constitutive of its value as a heritage site.

One of the generated images was a black and white photograph, which we
assume is in response to the specific mention of dates (1882/1884) in
the prompt potentially directing the colour scheme. The tonality, frame,
and context of H-M image) are closest to photographs of the late
nineteenth and early twentieth century archaeological surveys.

The subject of the M-M image is notionally closer to the original in
terms of morphology, materials, and a focus on roof form. The foreground
landscape echoes the materiality of the subject, while the background
reproduces vegetation and tonality often depicted in images of the
African savanna. The tight framing of the structure in the photograph
and the difficulty in assigning a sense of scale mimics the original
image, but the central difference between the two is in framing the
subject, which shifts the emphasis from the monumental in the original
to something more vernacular in the M-M image.

\hypertarget{anamorphosis-and-heteroscopia-two-properties-of-the-machinic-gaze}{%
\section{Anamorphosis and Heteroscopia: Two Properties of the Machinic
Gaze}\label{anamorphosis-and-heteroscopia-two-properties-of-the-machinic-gaze}}

The algorithmic reading and synthesis of the five UNESCO World Heritage
Sites offers an interesting counterpoint to UNESCO's own textual
descriptions. All five of the images, when read via BLIP-2, focused on
the descriptions of form, scale, material, and composition, erasing any
sense of aesthetic judgement or valuation and instead generating
descriptions for precision and conciseness with varying levels of
accuracy. For instance, while the caption generated for the historic
centre of Sanaá accurately identified `an aerial view of the old city of
Yemen', the caption generated for the photograph of the Old Towns of
Djenné was `a large sand castle with people walking in front of it'
while the photograph of the Tombs of Buganda Kings at Kasubi was `the
roof is made out of straw'. The misrecognition of the Great Mosque of
Djenné as a sandcastle reflects perhaps most clearly the distortion
introduced by a machinic reading of this kind. However, even the
simplification of the Kasubi tombs to essentially an image of a roof
allows us to reflect upon our own interpretation of the five images as
sites of globally recognised heritage. The second layer of algorithmic
reading of the image, via Google Vision's API followed a mathematical
extraction based on probabilistic interpretation. In each of the images,
elements such as \textquotesingle sky,' `grass,' and `building center,'
were identified, alongside other identifying descriptors such as
`medieval architecture', `arch,' `archaeological site', but also
specific descriptions such as `Classical architecture', or `Byzantine
architecture'. Occasionally seemingly contradictory descriptors would be
generated for the same image, once again illustrating the slippage
between image, and meaning in the absence of a referent informing the
machinic gaze.

To make sense of the specificity of machinic gaze we borrow two terms
from media analysis that have special relevance to current machine
learning techniques of image consumption and production: anamorphosis
and heterotopia. For Lacan (1977), anamorphosis refers to the oblique
angle that an image requires of a human subject to interpret its
representation. In our treatment here, we argue the machine produces
--~to varying degrees -- skewed perspectives and alternative
identifications that in turn pose the reflexive question of how
culturally settled ways of seeing -- as tourist, as heritage expert and
so on -- are themselves constituted. A mosque is interpreted as a `giant
sand castle' and image synthesis then renders this as an artefact that
appears to violate connotations of the original object: as something to
be gazed upon with reverence, to be preserved, and so on. This property
is present most obviously as a form of technical error that needs to be
corrected, either in the machine or in the specific instructions or
prompts given to it. However, the error is at the same itself generative
of questions as to how and why we perceive it \emph{as} an error -- what
in other words conditions our own acts of seeing and judging. Broadly
relevant to how we perceive the outputs of generative AI in any context,
in the field of heritage it queries, almost in the form of an implicitly
theoretical demand, how we attribute aesthetic value to certain objects
over others.

In concrete terms, the machinic gaze is partly anamorphic with respect
to human modes of perception due to the methods employed in their
training. Images are processed iteratively, first with coarse filters
that aim to identify, for example, horizontal and vertical lines, then
with finer filters that progressively distinguish more subtle gradations
in form and colour. Buildings -- as relatively geometrically regular
objects of a certain scale -- are likely to be seen as alike, regardless
of functional distinctions between, for example, a place of worship and
a playful structure designed to imitate a castle. Such distinctions, if
they feature at all, depend in turn upon the relative mass of images and
labels in the training set. Hence the apparent confusion between a
certain type of mosque and a sandcastle reflects the proportionate mass
of labelled images of Djenné, relative to other sites -- and the
corresponding value attributed, in the human (tourist and expert) gaze,
to that site. To `correct' this error would involve different practices
of touristic attention (or modified weightings of the training data), to
better `align' this vision with human expectation. Conversely, it is
precisely this orthogonal or anamorphic perspective that in turn
reflects upon existing practices of human observation and perspective -
the privileging of certain sites over others, the concentration of
canonical representations of `mosques' and `castles' and re-projection
of localised settings into the global imaginary of tourism and heritage.

As our explorations also show, the visual machines of Stable Diffusion
and Midjourney -- considered as an algorithm, a neural network, or an AI
system -- already `see' as a multiplicity. To describe this, we make use
of the concept of \emph{heteroscopia,} a term coined by Jaireth (2000).
Jaireth gives heteroscopia two meanings: the first refers to a general
scopic regime or visual culture of a historical period, while the second
refers to the ways a given image may incorporate or reference other
images, and so be more or less heteroscopic. Our own use adapts
Jaireth's concept to the context of computer vision. In these systems,
all image outputs are heteroscopic in Jaireth's second sense: they come
from nowhere other than an archive of existing images and are
necessarily and only images constituted by other images. But in this
machinic context we argue for two other meanings of the term. First,
`hetero' signifies something `other' -- an inhuman `gaze' that is
sometimes banal, sometimes nonsensical, sometimes sublime, but always
holding a fundamentally different relationship to the constellation of
object-image-text. The technical act of ``diffusion'' in models like
MidJourney and Stable Diffusion involves a twin process of adding and
subtracting noise to a large corpus of images to learn to discriminate
forms, styles, and colour compositions (Croitoru et al., 2023). Models
are trained to compose images, in other words, solely via a `universe'
of other images and their accompanying labels. They are heteroscopic in
the sense that no human could ever work from a written demand to an
image in a comparable way.

The second meaning of heteroscopia constitutes part of our claim for the
need for human interpretation of machine-generated images. Heteroscopia
in this sense refers to multiplicity of `gazes' these systems make
available in response to prompts. This is not simply a case, we argue,
of prompts containing stylistic cues that direct the system to produce
images concordant with, for instance, photorealism, surrealism,
pointillism, or some other aesthetic effect. In addition, these systems
build networks between prompt terms that unveil different -- and
sometimes, to human eyes, quite novel --~imagistic representations that
require a frame of interpretation. In the context of the heritage image,
we need a language that describes how, for instance, efforts to
reproduce the old towns of Djenné construct instead whimsical
sandcastles that impossibly dwarf human characters in the foreground. No
existing heritage taxonomic overlay can quite work to make sense of
these creations, and even existing artistic nomenclature would struggle
to `locate' these examples of machinic heteroscopia. This step of
algorithmic reading, which is devoid of the human `expert' or the
`tourist' gaze that relies on a constant referent to ideas of heritage
value but instead focuses purely on elements of the image reveals the
extent of meaning we implicitly attach to images of heritage sites.
Deploying the machinic gaze towards heritage photographs allows us to
occupy a position of tourist or expert or in some cases both, but in
each case, we are able to reflect upon the presumed author / generator
of the image. On the other hand, multiple historic and visual referents
are embedded within each of the five UNESCO site descriptions. Read
alongside the description, the image of the Djenné is inscribed with
multiple aesthetic judgements and associated ideas of heritage `value.'

\hypertarget{heritage-and-the-machinic-gaze}{%
\section{Heritage and the Machinic
Gaze}\label{heritage-and-the-machinic-gaze}}

The privileging of the visuality and aesthetics of heritage sites in
UNESCO is, we argue, distorted, and refracted through the machinic gaze,
and through the operations we identify as anamorphosis and heteroscopia.
In highlighting elements of both similarity and difference, through
visual representation, the fetishisation of architectural and artistic
form, ornamentation, and material can be examined through both sets of
images. In the first set, where the human textual description is used as
visual description / reinscription, we observe a greater degree of
diversity in both subject and framing, but consistent in the images
produced is a privileging of a certain kind of aesthetic that aligns to
the idea of heritage value being inscribed and prescribed visually. In
the second set, which is produced through a machinic reading and
resynthesis, even though the subject of the image shifts substantially,
the framing does not.

We argue that heteroscopia and anamorphosis help to cluster and
aggregate these features into refracted and concentrated delineations
that otherwise exist as more diffused tendencies or proclivities: how
the tourist and the expert sees. These tendencies appear more or less
evident across two of the site / model / prompt combinations. Sana'a (with Midjourney) is reproduced through something
like the tourist gaze -- imagined at a distance, with saturated colour ––
 while outputs prompted by Kasubi (with SDXL) prompts appear closer to an expert's view -- muted palette, with the photographic subject brought to the foreground. 
Djenné shares elements of both, but veers into alternative registers of the cinematic and
fantastical. In calling for such interpretations, the machine here acts
to bring these gazes themselves into focus. And with the act of
interpretation itself, we move invariably away from attention to purely
quantitative variances -- inherent in the very mechanisms by which
machine learning techniques aim to approximate a training set -- to
emphasise instead a process of human judgement and critique.

We conclude on a speculative note about the effects of this process.
MacCannell's (2001) critique of Urry's concept of the tourist gaze
stresses the reflexive character of the gaze left out in treatments that
discuss solely its dispositional, surveillant and outward looking
aspects. In Lacan's treatment of the gaze, which MacCannell draws upon,
its significance is the drawing back in of the viewing subject into the
picture or tableau (Lacan 2007). It is the subject who, alongside the
image under apprehension, at a critical moment perceives themselves as
being also observed, as an object that appears in the eyes of others.
The emergence of computer vision, machine learning and generative AI
exacerbates this reflexive moment. The human gaze -- especially in its
tourist or heritage genres -- becomes aware of itself in its
particularity, as a thing both distinctive and available in turn as
object for consumption by other viewers. The combined operation of
heteroscopia and anamorphosis here performs a kind of double act with
respect to human ways of seeing. Firstly, it points to the impossibility
of any privileged and original specular \emph{morphosis}, or
authoritative gaze. Secondly and conversely, it points to the
possibility of any given form of gaze becoming itself treated as
quotable reference material, and in that process, also becoming
objectified.

If, as Urry and Larsen (2011) and Sterling (2016) have argued, heritage
often preoccupies itself with objects in the form of a cliche or
ideology, the machinic gaze helps to make this ideology be seen
\emph{as} ideological: as an endowment of qualities worthy of respect
and preservation from imminent and long-term danger. Even as it is tied
up within its own networks of value and extraction, the machinic gaze
then serves to frame and comment upon how these ways of seeing
themselves structure and condition heritage objects and sites. It is not
so much that mechanical reproductions show what was previously invisible
as they~incorporate the human gazing figure within the frame -- if not
literally, then always indirectly -- making more conspicuous its varied
ideological character. In this sense computational vision seems to
contradict Urry and Larsen's claim about the earlier technology of
photography, which showed `miniature slices of reality, without
revealing its constructed nature or its ideological content' (Urry and
Larsen, 2011, p. 168). However, this may also be a temporary effect of
its novelty, and as this form of machinic gaze becomes less an object of
intense interest itself, becoming in its own way an invisible tourist of
human culture, its current reflexive character may dissolve into
background. Our own analysis invites sustained attention to a novel form
of seeing that in varied ways has us firmly in view.

\hypertarget{references}{%
\section{References}\label{references}}

Azar M, Cox G and Impett L (2021) Introduction: ways of machine seeing.
\emph{AI \& Society}: 1--12.

Barauah A (2017) \emph{Travel Imagery in the age of Instagram: An
ethnography of travel influencers and the ``online tourist gaze''}.
Master's Thesis. Department of Anthropology and Sociology SOAS,
University of London, London.

Benjamin W (1986) \emph{Illuminations}. New York: Random House.

Brown NE, Liuzza C and Meskell L (2019) The Politics of Peril: UNESCO's
List of World Heritage in Danger. \emph{Journal of Field Archaeology}
44(5): 287--303.

Buolamwini J and Gebru T (2018) Gender shades: Intersectional accuracy
disparities in commercial gender classification. In: \emph{Conference on
fairness, accountability and transparency}, 2018, pp. 77--91. PMLR.

Campt TM (2017) \emph{Listening to Images}. Durham, NC: Duke University
Press.

Crary J (1990) \emph{Techniques of the Observer}. Cambridge, MA: MIT
Press.

Croitoru F-A, Hondru V, Ionescu RT, et al. (2023) Diffusion models in
vision: A survey. \emph{IEEE Transactions on Pattern Analysis and
Machine Intelligence}. IEEE. Epub ahead of print 2023.

De Cesari C (2013) Thinking through heritage regimes. In: Bendix R,
Eggert A, and Peselmann A (eds) \emph{Heritage Regimes and the State}.
Göttingen: Universitätsverlag Göttingen, pp. 399--413.

Denicolai L (2021) The robotical gaze: A hypothesis of visual production
using technological image/thinking. \emph{MediArXiv preprint
https://mediarxiv.org/undh5/}. Epub ahead of print 2021.

Dicks B (2000) Encoding and Decoding the People: Circuits of
Communication at a Local Heritage Museum. \emph{European Journal of
Communication} 15(1): 61--78.

Du Camp M (1852) \emph{Egypte, Nubie, Palestine et Syrie: Dessins
Photographiques Recueillis de 1849 à 1851}. London: Gide et Baudry.

Jaireth S (2000) To see and be seen: the heteroscopia of Hindi film
posters. \emph{Continuum} 14(2): 201--214.

Kittler F (2010) \emph{Optical Media}. London: Polity.

Lacan J (1998) \emph{Book XI: The Four Fundamental Concepts of
Psychoanalysis} (tran. A Sheridan). New York: WW Norton \& Company.

Li J, Li D, Savarese S, et al. (2023) Blip-2: Bootstrapping
language-image pre-training with frozen image encoders and large
language models. \emph{arXiv preprint arXiv:2301.12597}. Epub ahead of
print 2023.

Luccioni AS, Akiki C, Mitchell M, et al. (2023) Stable bias: Analyzing
societal representations in diffusion models. \emph{arXiv preprint
arXiv:2303.11408}. Epub ahead of print 2023.

MacCannell D (2001) Tourist agency. \emph{Tourist studies} 1(1): 23--37.

MacCannell D (2013) \emph{The Tourist: A New Theory of the Leisure
Class}. Berkeley, CA: University of California Press.

Mackenzie A and Munster A (2019) Platform seeing: Image ensembles and
their invisualities. \emph{Theory, Culture \& Society} 36(5): 3--22.

McGowan T (2018) Cinema after Lacan. In: Mukherjee A (ed.) \emph{After
Lacan: Literature, Theory, and Psychoanalysis in the 21st Century}.
Cambridge: Cambridge University Press, pp. 115--28.

Midjourney (2022) We're officially moving to open-beta! Join now at
https://discord.gg/midjourney. **Please read our directions carefully**
or check out our detailed how-to guides here:
https://midjourney.gitbook.io/docs. Most importantly, have fun! In:
\emph{Twitter}. Tweet. Available at:
\url{https://twitter.com/midjourney/status/1547108864788553729}{\ul{https://twitter.com/midjourney/status/1547108864788553729}}
(accessed 5 October 2023).

Moshenska G (2013) The archaeological gaze. In: González-Ruibal A (ed.)
\emph{Reclaiming Archaeology: Beyond the Tropes of Modernity}. London:
Routledge, pp. 211--19.

Mostaque E (2022) Stable Diffusion Public Release. Available at:
\url{https://stability.ai/blog/stable-diffusion-public-release}{\ul{https://stability.ai/blog/stable-diffusion-public-release}}
(accessed 10 April 2022).

Mulvey L (2013) Visual pleasure and narrative cinema. In: Penley C (ed.)
\emph{Feminism and Film Theory}. London: Routledge, pp. 57--68.

Offert F and Phan T (2022) A sign that spells: DALL-E 2, invisual images
and the racial politics of feature space. \emph{arXiv preprint
arXiv:2211.06323}. Epub ahead of print 2022.

Ogden R (2021) Instagram photography of Havana: Nostalgia, digital
imperialism and the tourist gaze. \emph{Bulletin of Hispanic Studies}
98(1): 87--108.

Oh Y (2022) Insta-gaze: Aesthetic representation and contested
transformation of Woljeong, South Korea. \emph{Tourism Geographies}
24(6--7): 1040--1060.

Parisi L (2019) The alien subject of AI. \emph{Subjectivity} 12: 27--48.

Parisi L (2021) Negative optics in vision machines. \emph{AI \& Society}
36: 1281--1293.

Ramesh A, Dhariwal P, Nichol A, et al. (2022) Hierarchical
text-conditional image generation with clip latents. \emph{arXiv
preprint arXiv:2204.06125}. Epub ahead of print 2022.

Salvaggio E (2022) How to Read an AI Image. In: \emph{Cybernetic
Forests}. Available at:
\url{https://cyberneticforests.substack.com/p/how-to-read-an-ai-image}{\ul{https://cyberneticforests.substack.com/p/how-to-read-an-ai-image}}
(accessed 10 July 2023).

Schmidt K (2023) Description. Available at:
\url{https://github.com/kaikalii/stable-diffusion-artists}{\ul{https://github.com/kaikalii/stable-diffusion-artists}}
(accessed 21 March 2023).

Schuhmann C, Beaumont R, Vencu R, et al. (2022) Laion-5b: An open
large-scale dataset for training next generation image-text models.
\emph{arXiv preprint arXiv:2210.08402}. Epub ahead of print 2022.

Smith L (2006) \emph{Uses of Heritage}. London: Routledge.

Sterling C (2016) Mundane myths: Heritage and the politics of the
photographic cliché. \emph{Public Archaeology} 15(2--3): 87--112.

Sterling C (2019) \emph{Heritage, Photography, and the Affective Past}.
London: Routledge.

UNESCO (2008) The Criteria for Selection. Available at:
\url{https://whc.unesco.org/en/criteria/}{\ul{https://whc.unesco.org/en/criteria/}}
(accessed 5 October 2023).

UNESCO (2023) World Heritage List. Available at:
\url{https://whc.unesco.org/en/list/}{\ul{https://whc.unesco.org/en/list/}}
(accessed 5 October 2023).

Urry J and Larsen J (2011) \emph{The Tourist Gaze 3.0}. London: Sage.

Waterton E (2009) Sights of sites: Picturing heritage, power and
exclusion. \emph{Journal of Heritage Tourism} 4(1): 37--56.

Watson S and Waterton E (eds) (2016) \emph{Culture, Heritage and
Representation: Perspectives on Visuality and the Past}. London:
Routledge.

Winter T (2006) Ruining the dream? The challenge of tourism at Angkor,
Cambodia. In: Meethan K, Anderson A, and Miles S (eds) \emph{Tourism
Consumption and Representation: Narratives of Place and Self}.
Wallingford, UK: CABI, pp. 46--66.

Zhang J, Huang J, Jin S, et al. (2023) Vision-language models for vision
tasks: A survey. \emph{arXiv preprint arXiv:2304.00685}. Epub ahead of
print 2023.

\end{multicols*}

\end{document}